\documentclass[12pt]{article}
\usepackage[xdvi]{graphicx}
\def\be{\begin{equation}}
\def\ee{\end{equation}}
\def\bea{\begin{eqnarray}}
\def\eea{\end{eqnarray}}

\def\pe2{p_E^2}

\makeatletter

\@addtoreset{equation}{section}
\makeatother

\begin{document}
\newcommand{\mpl}{M_{\mathrm{Pl}}}
\setlength{\baselineskip}{18pt}
\begin{titlepage}
\begin{flushright}
KOBE-TH-06-03 \\
ROMA-1430/06 
\end{flushright}

\vspace{1.0cm}

\begin{center}
{\Large\bf Six Dimensional Gauge-Higgs Unification \\
[3mm]
with an Extra Space $S^2$ and the Hierarchy Problem} 
\end{center}

\vspace{10mm}

{\Large
\begin{center}
C. S. Lim$^{(a)}$
\footnote{e-mail: lim@kobe-u.ac.jp}, 
Nobuhito Maru$^{(b)}$ \footnote{e-mail: 
Nobuhito.Maru@roma1.infn.it} 
and K. Hasegawa$^{(c)}$
\footnote{e-mail: hasegawa@phys.ualberta.ca}, 

\end{center}}
\vspace{5mm}
\begin{center} 
$^{(a)}${\it Department of Physics, Kobe University, \\
Rokkodai, Nada, Kobe 657-8501, Japan} \\
[3mm]
$^{(b)}${\it Dipartimento di Fisica, 
Universit\`a di Roma "La Sapienza" \\
and INFN, Sezione di Roma, 
P.le Aldo Moro 2, I-00185 Roma, Italy} \\  
[3mm]
$^{(c)}${\it Department of Physics, 
University of Alberta, \\
Edmonton, Alberta, Canada} 
\end{center}
%
%
\vspace{1cm}
\centerline{\large\bf Abstract}
\vspace{0.5cm}
We calculate one-loop radiative correction to the mass of Higgs identified with 
the extra space components of the gauge field in a six dimensional massive scalar QED compactified on a two-sphere. 
The radiatively induced Higgs mass is explicitly shown to be finite for arbitrary bulk scalar mass $M$.  
Furthermore, the remaining finite part also turns out to vanish, at least for the case of small $M$, thus suggesting that the radiatively induced Higgs mass exactly vanishes, in general. The non-zero ``Kaluza-Klein" modes in the gauge sector are  argued to have a Higgs-like mechanism and quantum mechanical $N=2$ supersymmetry, while the Higgs zero modes, as supersymmetric states, have a close relation with monopole configuration.

\end{titlepage}

%
%
\section{Introduction} 

It is well-known that A. Einstein in later stage of his life attempted to unify at that time known two interactions, i.e. gravity and electromagnetic interactions, both being mediated by bosonic particles with spin $s=2$ and $1$. 
Nowadays we know there is another kind of interaction, Higgs interaction, and  
it may be natural to ask whether the unification of gauge and Higgs interactions, which are also mediated by bosonic particles whose spins differ by 1 unit, is possible. Such ``gauge-Higgs unification" is realized in a framework of higher dimensional gauge theory where the extra-space components of the gauge field are identified with Higgs fields \cite{FM, H}.    

The gauge-Higgs unification is one of the attractive scenarios 
beyond the Standard Model, since it has a possibility to solve the longstanding problems of the Standard Model related to the Higgs sector. For instance, in the scenario interactions are basically governed by gauge principle and there should  be no arbitrary parameters in the theory. It is also possible that gauge symmetry is dynamically broken by the VEV of the extra-space component of the gauge field, 
i.e. ``Hosotani mechanism" may be operative \cite{H}.  

Furthermore, more recently the scenario has attracted a revived interest as a  possible interesting solution to the hierarchy problem \cite{HIL, gaugehiggs1, ABQ, gaugehiggs2, warpgh, ghfiniteT, GIQ, MY}. 
Especially the hierarchy problem at the quantum level, so-called  
the problem of quadratic divergence in the radiatively induced Higgs mass, has been shown to be solved without invoking to supersymmetry. The reason is simply that, since the Higgs fields are identified with the 
extra space components of the gauge field, a local operator responsible for the Higgs mass-squared is strictly forbidden by the higher dimensional local gauge  symmetry. Thus we expect no UV-divergent quantum corrections to the Higgs mass. 

In fact, an explicit calculation shows that the radiative correction to the Higgs mass is rendered to be finite, once all ``Kaluza-Klein" modes are summed up in the intermediate state \cite{HIL}. Actually, however, the calculation tells us that the finite mass is non-vanishing, roughly of the order of 
$1/R$ ($R$: the size of the extra space).   
Hence, to solve the hierarchy problem, the size of the extra space $R$ should be 
roughly of the order $1 (TeV^{-1})$, unless the Higgs mass is exponentially suppressed by a factor $e^{-RM}$ ($M:$ the bulk mass of the matter field). 

At the first glance, this result of non-vanishing Higgs mass seems to contradict with the above statement that the local operator for the Higgs mass is strictly forbidden. What really happens is that the effective potential as the function of Wilson-loop is radiatively induces  \cite{H}. The Wilson loop is of course gauge invariant non-local operator without any derivatives for the gauge field, thus providing an operator for the  Higgs mass, which is free from 
any UV-divergence. 

The finiteness of the Higgs mass was investigated 
in the five dimensional (5D) QED compactified on $S^1$ 
\cite{HIL} and 6D $U(3) \times U(3)$ gauge theory 
with toroidal compactifications \cite{ABQ}. 
The finiteness was also discussed in the model with orbifold compactification 
and branes \cite{GIQ}. 
Recently, the finiteness of Higgs mass at two-loop level 
has been explicitly confirmed in 5D QED compactified on $S^1$ \cite{MY}. 
Even in the ``Gravity-Gauge-Higgs unification" scenario, where all interactions mediated by bosonic particles with all possible spins are unified in a framework of Kaluza-Klein type higher dimensional gravity theory, the same argument holds true and 
the Higgs mass at one-loop level was shown to be finite \cite{HLM}. 

There have been continuous attempts \cite{gaugehiggs2, warpgh} to construct realistic (beyond the standard) models based on the gauge-Higgs unification scenario and ``orbifolding" \cite{kawamura}, which helps to achieve chiral gauge theories and to break gauge symmetry by non-trivial $Z_{2}$ parity assignment.  
It is interesting to note that the stimulating scenarios of ``dimensional deconstruction" \cite{G} and ``higgsless models" \cite{higgsless} have some similarities to that of gauge-Higgs unification; The dimensional deconstruction may be regarded as a 5D gauge theory where the extra dimension is latticized. 
The sector of non-zero Kaluza-Klein modes of the higgsless models might be understood to be what we obtain by taking a ``unitary gauge" in the system of 4D gauge and scalar fields in the gauge-Higgs unification scenario.

So far, the finiteness of Higgs mass in the gauge-Higgs unification  
has been studied only in the limited types of compactifications, 
i,e, on a circle, torus, or orbifold.     
Therefore, it is worth while to check 
whether the Higgs mass is finite for other types of compactifications, such as 
higher dimensional sphere $S^{N} \ (N \geq 2)$. 

In this context, it should be stressed that the Wilson-loop is non-trivial for the case of $S^{1}$, even if the field strength vanishes everywhere, just because the circle is non-simply-connected space and the Wilson line cannot be shrunk into a point; a phenomenon similar to A-B effect is responsible for the Higgs mass. Let us note that 4-dimensional gauge field does not acquire any quantum correction to its mass, as the Wilson loop is trivial for the 4-dimensional Minkowski space. 

Thus it will be natural to ask what happens if the topology of the compact extra space is of different type. We may naively expect that the Wilson-loop becomes trivial for the case of simply-connected space, such as $S^{N} \ (N \geq 2)$, and therefore the quantum correction to the Higgs mass, not only is finite, but also exactly vanishes. Let us note if this is confirmed the size of the extra space needs not to be of the order $1 (TeV^{-1})$, and small extra dimensions, such as of the order of Planck length may be allowed, avoiding the hierarchy problem at the same time (without relying on a large bulk mass $M$).   
 
From such motivation, in this paper we calculate the one-loop correction to the Higgs mass in a six dimensional massive scalar QED compactified on two-sphere $S^2$. We confirm by an explicit calculation that the Higgs mass is finite for arbitrary bulk mass $M$ of the scalar field. In addition, we also demonstrate that actually the radiatively induced Higgs mass exactly vanishes, at least for the case of small $M$. We have not shown that the Higgs mass vanishes for arbitrary $M$, though we expect it is the case. (In \cite{HIL}, an earlier attempt was made to calculate the radiative correction in a too simplified model with only extra space $S^{2}$, ignoring the 4D Minkowski space-time.)  

We also argue the non-zero ``Kaluza-Klein" modes in the gauge sector possess a Higgs-like mechanism and quantum mechanical $N=2$ supersymmetry, as was discussed in \cite{LNSS}, while the Higgs zero modes, as supersymmetric states, have a close relation with monopole configuration. The 4D kinetic term for the Higgs zero modes turn out to be not normalizable and we argue how we should interpret this result 
in the context of the hierarchy problem.

\setcounter{equation}{0}
\section{Higgs Mass in 6D Scalar QED on $S^2$}
\subsection{Action and 4D effective Lagrangian}
Here we consider a six dimensional massive scalar QED compactified on $S^2$ 
and calculate the radiative correction to the mass-squared of the Higgs, which are identified with the extra space components of the 6D gauge field. 
The reason to consider a scalar QED, instead of ordinary QED with a fermion, is just for the sake of technical simplicity. Basically, we can choose any models, as long as they have higher dimensional gauge invariance. The scalar field is assumed to have a 6D ``bulk mass" $M$. The metric  $g_{MN}$ for the space-time $M^{4} \times S^{2}$ is given by a line element  
\bea
ds^2 = g_{MN}dx^{M} dx^{N} = \eta_{\mu\nu} dx^\mu dx^\nu 
- R^2 d^2 \theta - R^2 \sin^2\theta d^2 \varphi,  
\label{metric}
\eea
where we take the metric in four dimensions to be mostly minus as 
$\eta_{\mu\nu}=(1,-1,-1,-1)$$(\mu, \nu =0,1,2,3)$, and the polar coordinates on the two-sphere with the radius $R$ are denoted as $(\theta, \varphi)$.  

The action is given by 
\bea
S &=& \int d^4x \int_0^\pi d\theta \int_0^{2\pi} d\varphi 
\sqrt{-g} \left[ 
\frac{1}{4}g^{MP}g^{NQ} F_{MN}F_{PQ} \right. \nonumber \\
&& \left. 
+ g^{MN}[(\partial_M + ie A_M) \Phi^*][ (\partial_N \phi-ieA_N) \Phi] 
-M^2\Phi^* \Phi
\right]~(M, N, P, Q = 0 \sim 3, 5,6)  \nonumber \\
\label{action}
\eea
where the first term is the kinetic term 
of the gauge field $A_M$, the second one is the kinetic term of 
a complex scalar field $\Phi$ with charge $e$ and the last one is its bulk mass term.   

This model has a higher dimensional $U(1)$ gauge symmetry 
\bea
A_M(x, \theta, \varphi) \to A_M(x, \theta, \varphi) 
+ \partial_M \xi(x, \theta, \varphi)~(\xi: {\rm transformation~parameter}),
\eea
especially, the fifth and the sixth components of this gauge transformation 
(the shift symmetry) $A_{\theta,\varphi} \to A_{\theta,\varphi} 
+ \partial_{\theta,\varphi} \xi$ forbid the local operator responsible for the Higgs mass-squared at the classical level. 

Therefore, we have to calculate 
the quantum corrections to the Higgs mass-squared, which is described by an effective 
action 
\bea
S_{H mass} = \int d^4x \int_0^\pi d\theta \int_0^{2\pi} d\varphi \ \sqrt{-g}\frac{1}{2}m_{H}^2 g^{ab} A_{a} A_{b} \ \ (a,b = \theta, \ \varphi), 
\label{higgsmass} 
\eea
which has only $S^{2}$ general coordinate invariance, not full 6D  
invariance. We have checked the (``Kaluza-Klein" zero mode of) 4-dimensional gauge field $A_{\mu}$ never gets a quantum correction to its mass, just because the one-loop calculation in 6D scalar QED reduces to that in 4D scalar QED with an arbitrary scalar mass.   
 
In order to calculate the Higgs mass $m_{H}$ in (\ref{higgsmass}), 
we should choose a suitable field configuration for $A_{\theta}, \ A_{\varphi}$, which has no contribution to the field strength $F_{\theta \varphi}$ but contributes to the action $S_{H mass}$. Here, for technical simplicity of calculation, we choose a 
configuration $A_M=(0, 0, 0, 0, A_{\theta}, 0)$ with $A_{\theta} 
= c \ \sin \theta \ (c:{\rm constant~background~field})$, and calculate the one-loop self-energy diagram of $A_{\theta}$ due to the bulk scalar exchange with zero external 4-momentum, where this background configuration of $A_{\theta}$ is inserted. 

Substituting $A_M = (0, 0, 0, 0, c \ \sin \ \theta, 0)$ in (\ref{higgsmass}) and performing the $\theta$ and $\varphi$ integration, we get a 4D effective lagrangian as the quadratic function of the constant background $c$  
\bea 
{\cal L}_{{\rm eff}} = -\frac{4\pi}{3} c^2 m^2_{H}. 
\eea
Thus, once we obtain the effective lagrangian ${\cal L}_{{\rm eff}}$ by the  explicit calculation of the self-energy diagram, the Higgs mass $m_{H}$ is given by a relation  
\bea 
m_{H}^{2} = - \frac{3}{4\pi c^{2}} \ {\cal L}_{{\rm eff}}.  
\label{effective}
\eea

\begin{figure}[h]
  \begin{center}
  \includegraphics[width=10.5cm]{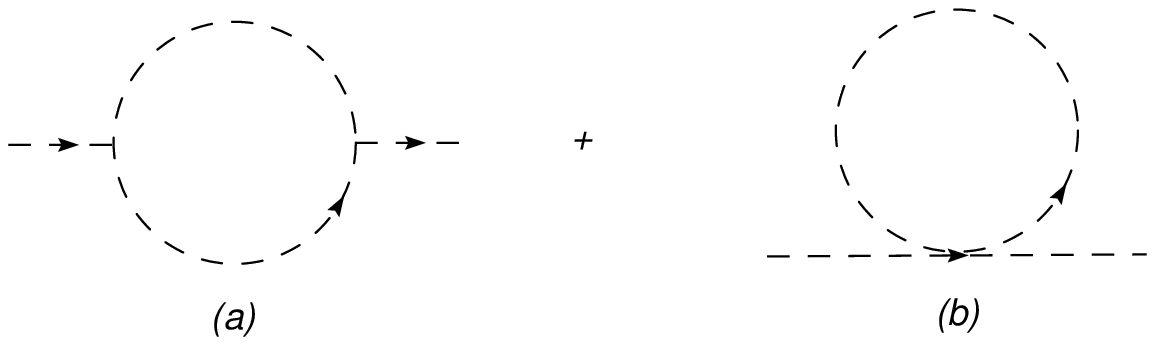}
  \put(-320,50){{\Large $A_\theta$}}
  \put(-258,95){{\LARGE $\hat{\phi}_{l+1,m}$}}
  \put(-255,35){{\LARGE $\hat{\phi}_{l,m}$}}
  \put(-180,50){{\Large $A_\theta$}}
  \put(-120,20){{\Large $A_\theta$}}
  \put(-65,50){{\LARGE $\hat{\phi}_{l,m}$}}
  \put(0,20){{\Large $A_\theta$}}
 \end{center}
\caption{Feynman diagrams contributing to the self-energy of $A_\theta$  
consist of (a) two three point vertices and (b) single four point vertex. 
The external lines denote $A_\theta$ and $\hat{\phi}_{l,m}$ running in the loop 
is the (canonically normalized) 4D scalar field with a Kaluza-Klein mode denoted by two integers $l,m$.} 
\label{diagram}
\end{figure}

The Feynman diagrams, contributing to the self-energy of $A_{\theta}$, are shown 
in Fig. \ref{diagram}. 

In order to calculate these diagrams, 
we need the 4D effective lagrangian for the bulk scalar field 
and to read off the necessary Feynman rules. 

Let the bulk scalar field $\Phi$ be expanded 
in terms of the spherical harmonic function $Y_l^m(\theta, \varphi)$, 
\bea
\Phi(x,\theta,\varphi) = \sum_{l=0}^\infty \sum_{m=-l}^l 
\phi_{l,m}(x)Y_l^m(\theta, \varphi)
\label{KKexp}
\eea
where $\phi_{l,m}(x)$ are 4D complex scalar fields. 
Then, the scalar part of the 4D lagrangian is obtained 
by putting (\ref{KKexp}) into (\ref{action}) and 
integrating with respect to $\theta, \phi$, under the configuration $A_{\theta} 
= c \ \sin \theta$; 
\bea
{\cal L}_S^{4D} &=& R^2 \int_0^\pi d\theta 
\int_0^{2\pi} d\varphi \sin \theta 
\left[
g^{MN}[(\partial_M + ie A_M) \Phi^*][ (\partial_N \phi-ieA_N) \Phi] 
-M^2 \Phi^* \Phi
\right] \nonumber \\
&=& R^2 \int_0^\pi d\theta 
\int_0^{2\pi} d\varphi \sin \theta 
\left[
\Phi^* \left( -\partial_\mu \partial^\mu + \frac{1}{R^2} \Delta_{S^2} 
- M^2 \right)\Phi \right. \nonumber \\
&& \left. \hspace*{3cm}-i\frac{ce}{R^2}
\left(
\Phi^* \sin \theta \partial_\theta \Phi 
-\Phi \sin \theta \partial_\theta \Phi^*
\right) 
-\frac{c^2 e^2}{R^2}\Phi^* \sin^2 \theta \Phi
\right] \nonumber \\
&=& 
\sum_{l=0}^\infty \sum_{m=-l}^l 
\left[
\hat{\phi}_{l,m}(x)^* 
\left( -\partial_\mu \partial^\mu -\frac{l(l+1)}{R^2} -M^2
\right)\hat{\phi}_{l,m}(x) \right. \nonumber \\
&& \left. -2 i \sqrt{4\pi}\frac{\hat{e}}{R}c (l+1)
\sqrt{\frac{(l+1)^2-m^2}{(2l+1)(2l+3)}}
(\hat{\phi}_{l+1,m}(x)^* \hat{\phi}_{l,m}(x) 
- \hat{\phi}_{l,m}(x)^* \hat{\phi}_{l+1,m}(x)) 
\right. \nonumber \\
&& \left. -4\pi\hat{e}^2 c^2 
\left\{
\frac{2(l^2+l-1+m^2)}{(2l-1)(2l+3)}\hat{\phi}_{l,m}^* \hat{\phi}_{l,m}(x) 
\right. \right. \nonumber \\
&& \left. \left. 
-\frac{\sqrt{((l+2)^2-m^2)((l+1)^2-m^2)}}{(2l+3)\sqrt{(2l+1)(2l+5)}} 
(\hat{\phi}_{l,m}(x)^* \hat{\phi}_{l+2,m}(x) 
+ \hat{\phi}_{l+2,m}(x)^* \hat{\phi}_{l,m}(x))
\right\}
\right] \nonumber \\ 
\label{4Deff}
\eea
where 
\bea
\Delta_{S^2} \equiv \frac{1}{\sin \theta}
\frac{\partial}{\partial \theta} \left( 
\sin\theta \frac{\partial}{\partial \theta} \right) 
+ \frac{1}{\sin^2\theta}\frac{\partial^2}{\partial \varphi^2}
\eea
in the second line is the laplacian on the two-sphere $S^2$. 
The third line is written in terms of 4D scalar field with 
mass dimension one, $\hat{\phi}_{l,m}(x) \equiv R \phi_{l,m}(x)$, 
and 4D gauge coupling $\hat{e} \equiv \frac{e}{\sqrt{4\pi}R}$. 
The detailed derivation of this 4D effective lagrangian of the scalar part 
(\ref{4Deff}) is described in appendix \ref{derive}. 

\subsection{One-loop calculation of the Higgs mass $m_H^{2}$}
\begin{figure}[ht]
  \begin{center}
  \includegraphics[width=11.5cm]{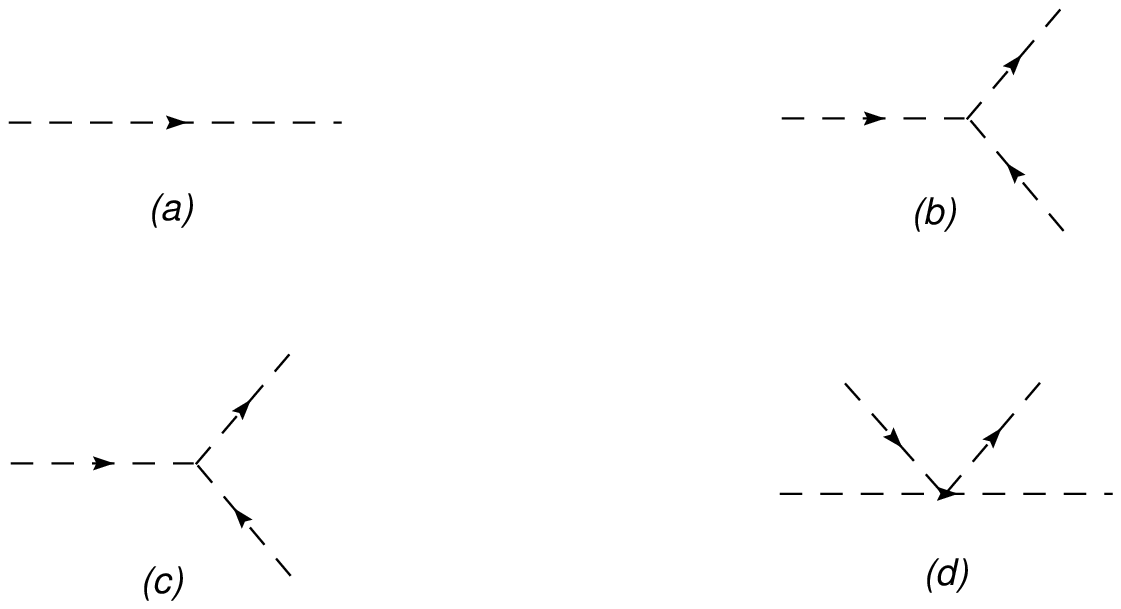}
  \put(-345,40){{\Large $A_\theta$}}
  \put(-240,00){{\Large $\hat{\phi}_{l+1,m}$}}
  \put(-285,155){{\Large $\hat{\phi}_{l,m}$}}
 \put(-120,140){{\Large $A_\theta$}}
  \put(-120,30){{\Large $A_\theta$}}
  \put(-240,80){{\Large $\hat{\phi}_{l,m}$}}
  \put(05,30){{\Large $A_\theta$}}
  \put(-100,75){{\Large $\hat{\phi}_{l,m}$}}
  \put(-20,75){{\Large $\hat{\phi}_{l,m}$}}
  \put(-10,105){{\Large $\hat{\phi}_{l,m}$}}
  \put(-10,180){{\Large $\hat{\phi}_{l+1,m}$}}
 \end{center} 
 \caption{Feynman rules necessary for the calculation of Fig. \ref{diagram}. 
 (a) denotes the propagator of the scalar fields, 
 (b) and (c) denote the three point vertices, 
 and (d) denotes the four point vertex. 
 Feynman rules irrelevant for Fig. \ref{diagram} are omitted 
 in this figure.} 
 \label{Feynman}
 \end{figure}

Now, we are in a position to calculate the one-loop induced Higgs mass-squared $m_{H}^{2}$. 
Feynman rules needed for self-energy diagrams in Fig. \ref{diagram} 
can be immediately read off from (\ref{4Deff}) as shown in Fig. \ref{Feynman}, where the propagator and each vertex are given by
\bea
(a) &=& \frac{i}{k^2-\frac{l(l+1)}{R^2}-M^2}, \\
(b) &=& 2\sqrt{4\pi}\frac{\hat{e}}{R}c(l+1)
\sqrt{\frac{(l+1)^2-m^2}{(2l+1)(2l+3)}}, \\
(c) &=& -2\sqrt{4\pi}\frac{\hat{e}}{R}c(l+1)
\sqrt{\frac{(l+1)^2-m^2}{(2l+1)(2l+3)}}, \\
(d) &=& -8\pi i \hat{e}^2 c^2 \frac{l^2+l-1+m^2}{(2l-1)(2l+3)} 
\eea 
where in (d) the vertices with the flip of $l$ with 2 units are irrelevant for the calculations of Fig. \ref{diagram}, and are simply neglected. 
Using these Feynman rules, it is straightforward to calculate 
the diagrams shown in Fig. \ref{diagram} to get the 4D effective lagrangian  
\bea
{\cal L}_{{\rm eff}} &=& \sum_{l,m}[(a) + (b)~{\rm in~Fig.~\ref{diagram}}] 
\nonumber \\
&=&(-i) c^2 \sum_{l,m} \int\frac{d^dk}{(2\pi)^d} \times\nonumber \\
&&\left[
\left( 2\sqrt{4\pi} \frac{\hat{e}}{R} \right)^2 \frac{(l+1)^2-m^2}{(2l+1)(2l+3)}
(l+1)^2 \frac{1}{\left[ k^2-\frac{(l+1)(l+2)}{R^2}-M^2 \right] 
\left[ k^2 -\frac{l(l+1)}{R^2} - M^2 \right]} \right. \nonumber \\
&&\left. \hspace*{4cm}+8\pi\hat{e}^2 \frac{l^2+l-1+m^2}{(2l-1)(2l+3)} 
\frac{1}{k^2-\frac{l(l+1)}{R^2}-M^2}
\right] \\
\label{1-loop}
&=& (-i)8\pi\hat{e}^2 c^2 \sum_{l=0}^\infty \int\frac{d^dk}{(2\pi)^d} 
\left[
\frac{2(l+1)^3}{3R^2} \frac{1}{\left[ k^2-\frac{(l+1)(l+2)}{R^2}-M^2 \right] 
\left[ k^2 -\frac{l(l+1)}{R^2} - M^2 \right]} \right. \nonumber \\
&& \left. \hspace*{4.5cm}+\frac{2l+1}{3}
\frac{1}{k^2-\frac{l(l+1)}{R^2}-M^2}
\right]
\label{1loop2}
\eea
where the dimensions of the momentum integration is analytically continued 
into $d$ dimensions in the second line in order to regularize the integral and the summation, in the spirit of dimensional regularization.  
In the last line, the sum with respect to $m$ is carried out. 
Noticing the relation
\bea
(l+1)^3 &=& \frac{1}{4}
\left[
(2l+3)l(l+1) +(2l+1)(l+1)(l+2) +(l+1)(l+2) -l(l+1)
\right] \nonumber \\
&=& \frac{R^2}{4}
\left[
-(2l+3)\left(k^2 -\frac{l(l+1)}{R^2} -M^2 \right) 
-(2l+1)\left(k^2 -\frac{(l+1)(l+2)}{R^2} -M^2 \right) 
\right. \nonumber \\
&& \left. +\left(k^2 -\frac{l(l+1)}{R^2} -M^2 \right) 
-\left(k^2 -\frac{(l+1)(l+2)}{R^2} -M^2 \right) 
+ 4(l+1)(k^2-M^2)
\right], \nonumber \\
\eea
(\ref{1loop2}) can be rewritten as follows. 
\bea
{\cal L}_{{\rm eff}} &=& (-i)8\pi\hat{e}^2 c^2 \sum_{l=0}^\infty 
\int\frac{d^dk}{(2\pi)^d} 
\left[
-\frac{1}{6}\left( \frac{2l+3}{k^2-\frac{(l+1)(l+2)}{R^2}-M^2} 
- \frac{2l+1}{k^2-\frac{l(l+1)}{R^2}-M^2} \right) \right. \nonumber \\
&& \left. +\frac{1}{6}\left( \frac{1}{k^2-\frac{(l+1)(l+2)}{R^2}-M^2} 
- \frac{1}{k^2-\frac{l(l+1)}{R^2}-M^2} \right) 
\right. \nonumber \\
&& \left. +\frac{2}{3}\frac{(l+1)(k^2-M^2)}
{[k^2-\frac{(l+1)(l+2)}{R^2}-M^2][k^2-\frac{l(l+1)}{R^2}-M^2]}
\right]  
\label{cancel} \\
&=&
(-i)8\pi\hat{e}^2 c^2 \sum_{l=0}^\infty 
\int\frac{d^dk}{(2\pi)^d} 
\frac{2}{3}\frac{(l+1)(k^2-M^2)}
{[k^2-\frac{(l+1)(l+2)}{R^2}-M^2][k^2-\frac{l(l+1)}{R^2}-M^2]}, 
\label{1loop3}
\eea
where one can easily check that the terms 
in the first and the second lines in (\ref{cancel}) exactly cancel out. 
\bea
&&-\frac{1}{6}\sum_{l=0}^\infty 
\left( \frac{2l+3}{k^2-\frac{(l+1)(l+2)}{R^2}-M^2} 
- \frac{2l+1}{k^2-\frac{l(l+1)}{R^2}-M^2} \right)  \nonumber \\
&&  +\frac{1}{6} \sum_{l=0}^\infty 
\left( \frac{1}{k^2-\frac{(l+1)(l+2)}{R^2}-M^2} 
- \frac{1}{k^2-\frac{l(l+1)}{R^2}-M^2} \right) \nonumber \\
&=&-\frac{1}{6} \left( \sum_{l=1}^\infty 
\frac{2l+1}{k^2-\frac{l(l+1)}{R^2}-M^2} 
- \sum_{l=1}^\infty \frac{2l+1}{k^2-\frac{l(l+1)}{R^2}-M^2} -\frac{1}{k^2-M^2} 
\right)  \nonumber \\
&&  +\frac{1}{6}\left( \sum_{l=1}^\infty 
\frac{1}{k^2-\frac{l(l+1)}{R^2}-M^2} 
- \sum_{l=1}^\infty \frac{1}{k^2-\frac{l(l+1)}{R^2}-M^2} -\frac{1}{k^2-M^2} 
\right) \nonumber \\
&=& 0. 
\eea
By use of Feynman parameter $t$, (\ref{1loop3}) can be expressed as 
\bea
{\cal L}_{{\rm eff}} &=& -\frac{16\pi}{3}\hat{e}^2 c^2 
\int_0^1 dt \sum_{l=1}^\infty \int\frac{d^dk}{(2\pi)^d} 
\frac{l(k^2+M^2)}{[k^2+M^2 +\frac{l^2+(2t-1)l}{R^2}]^2},  
\eea
where the Wick rotation is carried out. 
It is convenient to rewrite this expression further 
as follows; 
\bea
{\cal L}_{{\rm eff}} &=& -\frac{16\pi}{3}\hat{e}^2 c^2 
\int_0^1 dt \sum_{l=1}^\infty \int\frac{d^dk}{(2\pi)^d} l
\left[
\frac{1}{k^2+M^2 +\frac{l^2+(2t-1)l}{R^2}} 
- \frac{(l^2+(2t-1)l)/R^2}{[k^2+M^2 +\frac{l^2+(2t-1)l}{R^2}]^2} 
\right] \nonumber \\
&=& -\frac{16\pi}{3}\hat{e}^2 c^2 
\int_0^1 dt \left. \left( 1+\frac{\partial}{\partial \alpha} \right) 
\right|_{\alpha=1} \frac{\partial}{\partial M^2 
}\sum_{l=1}^\infty l \int\frac{d^dk}{(2\pi)^d} 
\ln \left( k^2 + M^2 +\alpha \frac{l^2+(2t-1)l}{R^2} \right),  
\nonumber \\
\label{derieffpot}
\eea
where $\alpha$ is a fictitious parameter. 
It is easy to see that this expression vanishes for ``de-compactification" limit, 
$R \ \to \ \infty$. In this limit, we may replace $\frac{l^2+(2t-1)l}{R^2}$ and $\frac{2l}{R^{2}}$ into $k_{5}^{2} + k_{6}^{2}$ and $dk_{5} dk_{6}$, respectively by use of extra space momenta $k_{5,6}$.  
Thus, in this limit (\ref{derieffpot}) reduces to 
\bea
{\cal L}_{{\rm eff}} &\to& -\frac{32\pi^{3}}{3}R^{2} \hat{e}^2 c^2 
\left. \left( 1+\frac{\partial}{\partial \alpha} \right) 
\right|_{\alpha=1} \frac{\partial}{\partial M^2 
} \int\frac{d^dk dk_{5}dk_{6}}{(2\pi)^{d+2}} 
\ln \left( k^2 + M^2 +\alpha (k_{5}^{2}+k_{6}^{2}) \right) \nonumber \\ 
&=& -\frac{32\pi^{3}}{3}R^{2} \hat{e}^2 c^2 
\left. \left( 1+\frac{\partial}{\partial \alpha} \right) 
\right|_{\alpha=1} \frac{\partial}{\partial M^2 
} \frac{1}{\alpha} \int\frac{d^dk dk_{5}dk_{6}}{(2\pi)^{d+2}} 
\ln \left( k^2 + M^2 + k_{5}^{2}+k_{6}^{2} \right) \nonumber \\ 
&=& 0. 
\eea
In the last step, the change of variables for $k_{5}$ and $k_{6}$ was made. 

(\ref{derieffpot}) is similar to the calculation of the Casimir energy. For the cases with odd-dimensional sphere as the extra space, the ``Kaluza-Klein" mode sum can be 
analytically performed. In fact, for the case of $S^{1}$, after the mode sum the remaining momentum integration was apparently super-convergent and the finiteness of $m_{H}$ was trivial \cite{HIL}. 
In our case with $S^{2}$, unfortunately the result of mode sum cannot be written by a simple analytic function, and the finiteness of $m_{H}$ is not trivial. 
In this paper, therefore, we take another approach. Namely, we first perform the 4D momentum integration by use of the dimensional regularization method. The result is written 
in terms of gamma function $\Gamma (- \epsilon)$ with a pole and the sum of  Riemann's zeta-functions $\zeta(N-2\epsilon) \ (N: \mbox{integer})$, coming from the remaining mode sum over $l$ ($\epsilon \equiv \frac{d-4}{2}$). Paying special attention to the fact that $\zeta(z)$ has a pole only at $z=1$ we expand the expression in terms of the power of $\epsilon$, and finally we take the limit of $\epsilon \to 0$. If we take $\epsilon \ \to \ 0$ from the beginning 
and replace the mode sum by zeta-functions $\zeta (N)$ later on, the result will be different. A simple example to show the validity of our method is given in appendix \ref{zeta}.     
   
We follow this approach. 
First by use of the dimensional regularization 
the momentum integration provides  
\be  
\int\frac{d^dk}{(2\pi)^d} 
\ln \left( k^2 + M^2 +\alpha \frac{l^2+(2t-1)l}{R^2} \right) 
= - \frac{1}{(4\pi)^{\frac{d}{2}}} \ \Gamma 
\left( -\frac{d}{2} \right) 
\left[\alpha \frac{l^2+(2t-1)l}{R^2} + M^{2} \right]^{\frac{d}{2}}, 
\ee
where $\Gamma(z)$ is the Gamma function. Thus we obtain 
\be 
{\cal L}_{{\rm eff}} = \frac{2d}{3}\frac{\hat{e}^2 c^2}{(4\pi)^{\frac{d}{2}-1}} 
\int_0^1 dt \ \Gamma\left( -\frac{d}{2} \right) \sum_{l=1}^\infty l 
\left( \frac{d}{2} \frac{l^2+(2t-1)l}{R^2} + M^2 \right) 
\left( \frac{l^2+(2t-1)l}{R^2} + M^2 \right)^{\frac{d}{2}-2}. 
\label{DR}
\ee
By use of the relation (\ref{effective}) we thus get the formula for the Higgs mass-squared  
\be
m^2_{H} = -\frac{2d\hat{e}^2}{(4\pi)^{\frac{d}{2}}} 
\int_0^1 dt \ \Gamma\left( -\frac{d}{2} \right) \sum_{l=1}^\infty l 
\left( \frac{d}{2} \frac{l^2+(2t-1)l}{R^2} + M^2 \right) 
\left( \frac{l^2+(2t-1)l}{R^2} + M^2 \right)^{\frac{d}{2}-2}. 
\label{mass}
\ee

\subsection{Finiteness of $m^2_{H}$}
Let us show that the Higgs mass-squared (\ref{mass}) actually is finite, thus solving the hierarchy problem of quadratic divergence.  

In (\ref{mass}), we may naively expect that the term $\left( \frac{l^2+(2t-1)l}{R^2} + M^2 \right)^{\frac{d}{2}-2}$ can be replaced by 1 in the limit of 
$\epsilon \ \to \ 0$. However, in order to properly take into account the effect of the pole of zeta function at $z = 1$, we have to Taylor-expand this factor in the inverse powers of $l$ as follows.  
\bea
m^2_{H} &=& \lim_{\epsilon \to 0}
 -\frac{(4+2\epsilon)^2 \hat{e}^2}{(4\pi)^{2+\epsilon}R^{2+2\epsilon}} 
\int_0^1dt \ \Gamma\left( -2-\epsilon \right) \times \nonumber \\
&&\sum_{l=1}^\infty l 
\left( l^2+(2t-1)l + \frac{(MR)^2}{2+\epsilon} \right) 
\left( l^2+(2t-1)l + M^2 R^2 \right)^\epsilon \nonumber \\
&=& \lim_{\epsilon \to 0}
 -\frac{(4+2\epsilon)^2 \hat{e}^2}{(4\pi)^{2+\epsilon}R^{2+2\epsilon}} 
\int_0^1dt \ \Gamma\left( -2-\epsilon \right) \times \nonumber \\
&&\sum_{l=1}^\infty  
\left[ l^{3+2\epsilon} +(2t-1)l^{2+2\epsilon} +\frac{(MR)^2}{2+\epsilon} 
l^{1+2\epsilon} 
\right]\left[ 1+\frac{2t-1}{l}+\frac{(MR)^2}{l^2} \right]^\epsilon \nonumber \\
&=& \lim_{\epsilon \to 0}
 -\frac{(4+2\epsilon)^2 \hat{e}^2}{(4\pi)^{2+\epsilon}R^{2+2\epsilon}} 
\int_0^1dt \ \Gamma\left( -2-\epsilon \right) \nonumber \\
&&\sum_{l=1}^\infty  
\left[ l^{3+2\epsilon} +(2t-1)l^{2+2\epsilon} 
+\frac{(MR)^2}{2+\epsilon} l^{1+2\epsilon} 
\right] \times \nonumber \\
&&\left[
1 + \epsilon \left( \frac{2t-1}{l} +\frac{(MR)^2}{l^2} \right) 
+ \frac{\epsilon(\epsilon-1)}{2!} \left( \frac{2t-1}{l}+\frac{(MR)^2}{l^2} 
\right)^2 \right. \nonumber \\
&& \left. + \frac{\epsilon(\epsilon-1)(\epsilon-2)}{3!} 
\left( \frac{2t-1}{l}+\frac{(MR)^2}{l^2} \right)^3 + \cdots 
\right],  
\label{expansion}
\eea
where $\epsilon \equiv \frac{d-4}{2}$. 

The possible divergent part of (\ref{expansion}) 
comes from the pole of the Gamma function 
$\Gamma(-2 - \epsilon)\sim -\frac{1}{2\epsilon}$. 
The pole is multiplied by the sum of zeta functions 
\bea 
\zeta (z) &=& \sum_{l=1}^{\infty} \ \frac{1}{l^{z}}.  
\eea
When we consider only the possible divergent part, 
in the Taylor expansion it seems that only the leading term 1 should be kept, since all other terms are multiplied by $\epsilon$. $\zeta(1-2\epsilon)$, however, 
has a pole of the first order 
($- \frac{1}{2\epsilon}$), and the $\epsilon$ in the expansion coefficients 
is canceled by the pole. Thus only the terms including $\zeta(1-2\epsilon)$ have 
nonvanishing contributions to the divergent part even in the limit $\epsilon \to 0$. 
Thus, by use of 
\bea
\int_0^1 dt (2t-1)^n = \left\{ 
\begin{array}{l}
0~(n:{\rm odd}) \\
\frac{1}{n+1}~(n:{\rm even})
\end{array}
\right.  
\eea
the term of ${\cal O} (1/\epsilon)$ is calculated to be 
\bea
(m^2_{H})_{{\rm div}} &=& 
\lim_{\epsilon \to 0}
 -\frac{(4+2\epsilon)^2 \hat{e}^2}{(4\pi)^{2+\epsilon}R^{2+2\epsilon}} 
\int_0^1dt \left( -\frac{1}{2\epsilon} \right) \times \nonumber \\
&&\left[
\zeta(-3-2\epsilon) +\frac{(MR)^2}{2+\epsilon} \zeta(-1-2\epsilon) \right. \nonumber \\
&& \left. + \epsilon \left\{ 
\frac{(MR)^4}{2+\epsilon} +\frac{(\epsilon-1)}{2!} 
\left( (2t-1)^2 \frac{(MR)^2}{2+\epsilon} 
+ 2(2t-1)^2(MR)^2  +(MR)^4  \right) \right. \right. \nonumber \\
&& \left. \left. +\frac{(\epsilon-1)(\epsilon-2)}{3!} 
\left( (2t-1)^4 +3(2t-1)^2(MR)^2 \right) \right. \right. \nonumber \\
&& \left. \left. +\frac{(\epsilon-1)(\epsilon-2)(\epsilon-3)}{4!}
(2t-1)^4 \right\}
\zeta(1-2\epsilon)
\right] \nonumber \\
&=& \lim_{\epsilon \to 0}
 -\frac{16\hat{e}^2}{(4\pi R)^2} 
\left( -\frac{1}{2\epsilon} \right) 
\left[ \zeta(-3) -\left( \frac{1}{2} \right) 
\left( \frac{2}{6} \right) \left( \frac{1}{5} \right)
+\left( \frac{1}{2} \right) 
\left( \frac{6}{24} \right) \left( \frac{1}{5}\right) \right. \nonumber \\
&& \left. +(MR)^2 \left\{ \frac{1}{2} \zeta(-1) 
+ \frac{1}{4}\left( \frac{1}{6} +\frac{2}{3} \right) 
- \left( \frac{1}{2} \right) \left( \frac{2}{6} \right) 
\right\} \right. \nonumber \\
&& \left. +(MR)^4 \left\{ \left(-\frac{1}{2} \right) \left( \frac{1}{2}\right) 
- \frac{1}{2}\left( -\frac{1}{2} \right) \right\}
\right] \nonumber \\
&=& 0, 
\label{div}
\eea
where we used $\zeta(1-2\epsilon) \simeq -\frac{1}{2\epsilon}$ and 
$\zeta(-3) = \frac{1}{120}, \zeta(-1)=-\frac{1}{12}$. 
We thus have found that the divergent term vanishes. 

\subsection{The finite part of $m_{H}^{2}$} 
In the previous subsection, 
we have confirmed that the Higgs mass is (at most) finite. 
In this subsection, 
we concentrate on the remaining finite part, and will see whether the finite part also vanishes or not. 
The finite contributions have two sources. One is the product of the pole of the Gamma function, $-\frac{1}{2\epsilon}$, with the ${\cal O}(\epsilon)$ terms in the factor which multiplies the Gamma function. Another source is the product of the poles of the Gamma function and the zeta function $\zeta(1-2\epsilon)$, $(-\frac{1}{2\epsilon})^{2}$, multiplied by the terms of ${\cal O}(\epsilon^{2})$.   
The finite part is thus calculated to be 
\bea
(m^2_{H})_{{\rm finite}} &=& \lim_{\epsilon \to 0}
 -\frac{\hat{e}^2}{\pi^{2}R^{2}} 
\int_0^1dt \ \Gamma\left( -2-\epsilon \right) \times \nonumber \\ 
&&\sum_{l=1}^\infty 
\left( l^{3+2\epsilon}+(2t-1)l^{2+2\epsilon} 
+ \frac{(MR)^2}{2+\epsilon} l^{1+2\epsilon} \right) \times \nonumber \\
&&
\left[ 1 + \sum_{n=1}^{\infty} \frac{\epsilon(\epsilon-1)\cdots(\epsilon-(n-1))}{n!}
\left(\frac{2t-1}{l} +\frac{(MR)^2}{l^2} \right)^n \right]
\nonumber \\
&=&\lim_{\epsilon \to 0}
 -\frac{\hat{e}^2}{\pi^{2}R^{2}}  
\int_0^1dt \left( -\frac{1}{2\epsilon} \right) \times \nonumber \\
&&\sum_{l=1}^\infty 
\left( l^{3+2\epsilon}+(2t-1)l^{2+2\epsilon} 
+ \frac{(MR)^2}{2+\epsilon} l^{1+2\epsilon} \right) \times \nonumber \\
&& \left[1+\sum_{n=1}^{\infty} \frac{\epsilon(\epsilon-1)\cdots(\epsilon-(n-1))}{n!}
\sum_{p=0}^n \frac{n!}{p!(n-p)!}
\left(\frac{2t-1}{l} \right)^{n-p}\left(\frac{(MR)^2}{l^2} \right)^p \right]
\nonumber \\
&=&
 \frac{\hat{e}^2}{2 \pi^2 R^2} 
\left[ \frac{1}{240} + \frac{1}{48}(MR)^{2} -\frac{1}{8}(MR)^{4}
\right. \nonumber \\
&& \left. + \int_0^1 dt \left[ (-2)\left\{ \zeta'(-3) 
+ \left( \frac{1}{2} \zeta'(-1) + \frac{1}{8} \zeta (-1) \right) (MR)^{2} \right\} \right. \right. \nonumber \\
&& \left. \left. +{\sum_{n=1}^{\infty}}' 
{\sum_{p=0}^n}' \frac{(-1)^{n-1}(n-1)!}{p!(n-p)!}
(2t-1)^{n-p}(MR)^{2p} \right. \right. \nonumber \\
&&\times \left. \left. 
\left( \zeta(n+p-3) + (2t-1) \zeta(n+p-2) 
+ \frac{(MR)^2}{2} \zeta(n+p-1) \right) \right] \right]
\label{massivefinite}
\eea
where ${\sum_{n=1}^{\infty}}' {\sum_{p=0}^n}' $ means that 
when $\zeta(1)$ appears in the sum it should be replaced by its 
finite part, i.e.    
\bea
\lim_{z \to 1}\left[\zeta(z) -\frac{1}{z-1} \right] = \gamma
\eea
where $\gamma$ is the Euler constant. 
The first line in the last equation in (\ref{massivefinite}) comes from 
the product of the poles of the Gamma and zeta functions 
accompanied by ${\cal O}(\epsilon^{2})$ terms, which happens only for the cases 
of $n+p=4,3,2$ in the expansion.  

First let us check whether this expression (\ref{massivefinite}) 
vanishes for the case of massless scalar $M = 0$ ($p=0$). 
In this case the expression reduces to 
\bea
(m^2_{H})_{{\rm finite}} &=& \frac{\hat{e}^2}{2 \pi^2 R^2} 
\left[ \frac{1}{240} -2\zeta'(-3) 
+ {\sum_{n=1}^{\infty}}' \frac{\zeta(2n-3)}{(2n-1)2n(2n+1)} \right] \nonumber \\
&=& \frac{\hat{e}^2}{2\pi^2 R^2} 
\left[ \frac{1}{240} -2\zeta'(-3) + \frac{\zeta(-1)}{6} + \frac{\gamma}{60} 
+ \sum_{n=1}^\infty \frac{\zeta(2n+1)}{(2n+3)(2n+4)(2n+5)} \right]. 
\nonumber 
\label{masslessfinite}
\\
\eea
Here let us note a useful mathematical relation shown in appendix \ref{formula}: 
\bea 
\sum_{n=1}^\infty \frac{\zeta(2n+1)}{(2n+3)(2n+4)(2n+5)} 
&=& \frac{1}{720}\left( 7 - 12 \gamma +1440 \zeta'(-3) \right) 
\eea
Then (\ref{masslessfinite}) is actually shown to vanish 
\bea
(m^2_{H})_{{\rm finite}} 
&=& \frac{\hat{e}^2}{2\pi^2 R^2} 
\left[ \frac{1}{240} -2\zeta'(-3) + \frac{\zeta(-1)}{6} + \frac{\gamma}{60} 
+\frac{1}{720}\left( 7 - 12 \gamma +1440 \zeta'(-3) \right) \right] \nonumber \\
&=& 
\frac{\hat{e}^2}{2\pi^2 R^2} 
\left[ \frac{1}{240} - \frac{1}{72} +\frac{7}{720} \right] = 0, 
\label{masslessfinite2}
\eea
where $\zeta(-1) = - \frac{1}{12}$. 

Next let us check whether $m_{H}^{2}$ vanishes even in the case of massive scalar field. Here we will focus on a specific case of small bulk mass $M \ll \frac{1}{R}$. Namely we calculate the ${\cal O} ((MR)^{2})$ term in (\ref{massivefinite}), which reads as 
\be 
(m^2_{H})_{{\rm finite}} 
= \frac{\hat{e}^2}{2\pi^2 R^2}(MR)^{2}  
\left[\frac{1}{48} - \zeta'(-1) + \frac{3}{4} \zeta(-1) 
- \frac{\gamma}{12} 
- \frac{1}{4} \sum_{n=1}^\infty \frac{\zeta(2n+1)}{(n+1)(2n+3)} \right] .  
\label{massivefinite2} 
\ee
Again using a useful mathematical relation which can be shown in a similar manner to that in appendix \ref{formula},  
\bea
\sum_{n=1}^\infty \frac{\zeta(2n+1)}{(n+1)(2n+3)} 
&=& - \frac{1}{3}\gamma - \frac{1}{6} -4 \zeta'(-1) ,  
\label{formula2}
\eea
$m_{H}^{2}$ turns out to vanish:  
\bea  
(m^2_{H})_{{\rm finite}} 
&=& \frac{\hat{e}^2}{2\pi^2 R^2}(MR)^{2}  
\left[\frac{1}{48} - \zeta'(-1) + \frac{3}{4} \zeta(-1) 
- \frac{\gamma}{12} 
+ \frac{\gamma}{12} + \frac{1}{24} + \zeta'(-1) \right]  \nonumber \\ 
&=& 0.   
\label{massivefinite3} 
\eea

Though we have not checked whether $m_{H}^{2}$ disappears for the case of 
general massive scalar QED, the obtained results strongly suggest that $m_{H}^{2}$ exactly vanishes for arbitrary 
$M$. These results are quite consistent with the physical understanding 
that the Wilson line on the two-sphere can be always shrunk to a point  
and therefore the finite Higgs mass is not generated at quantum level.  

\section{The Higgs Zero-Modes and Quantum Mechanical Supersymmetry for Non-Zero Modes}  

Having shown that quantum correction to the Higgs mass-squared $m_{H}^{2}$ exactly vanishes, 
we should now discuss the 4D mass spectra of $A_{\theta}, \ A_{\varphi}$, in order to identify the Higgs field with some Kaluza-Klein modes of these fields. 

Concerning the non-zero modes with non-vanishing 4D masses, we pay attention to the claim that, in general, higher dimensional gauge theories have $N=2$ quantum 
mechanical supersymmetry \cite{LNSS}. According to this argument, in each of 
the non-zero modes, a Higgs-like mechanism is operative, where (some of) the extra space components of the gauge field play the role of would-be Nambu-Goldstone (N-G) boson, and the eigenfunctions of differential operators to fix 4D mass-squared for $A_{\mu}$ and the extra space component of gauge field form a super-multiplet. 
The set of differential operators can be written in a form of supersymmetric Hamiltonian. The zero modes, on the other hand, do not form a super-multiplet, and are interpreted as the isolated supersymmetric states.  

We demonstrate below that our 6D QED really has such property. What we discuss is the kinetic term of the gauge field $A_{M}$ 
\be 
S_{gauge} = \int d^4x \int_0^\pi d\theta \int_0^{2\pi} d\varphi 
\sqrt{-g} \frac{1}{4}g^{MP}g^{NQ} F_{MN}F_{PQ}.  
\label{zeromode1} 
\ee
In order to make our argument transparent, we assume here that all fields $A_{\mu}, A_{\theta}, A_{\varphi}$ are $\varphi$-independent. Namely we focus on the modes with 
zero ``magnetic quantum number" $m$. Then the mixing terms between $A_{\varphi}$ and 
$A_{\mu}, \ A_{\theta}$ in (\ref{zeromode1}) disappears, and the Higgs-like 
mechanism should be operative in the $(R A_{\mu}, A_{\theta})$ multiplet, where the non-zero modes of $A_{\theta}$ behave as would-be N-G bosons. 
It is easy to know that the differential operators to fix 4D mass-squared for $A_{\mu}$ is just $\Delta_{S^{2}}$, and the field can be expanded in terms of Legendre polynomials $P_{l}(\cos \theta)$: 
\be 
A_{\mu} = \sum_{l=0}^{\infty} \frac{A_{\mu}^{(l)}(x)}{R} f_{l}(\theta), \ \ f_{l}(\theta) = \sqrt{\frac{2l+1}{4\pi}} P_{l}(\cos \theta).     
\ee
$f_{l}(\theta)$ are eigenfunctions of the eigenvalue equation  
\be 
   - \frac{1}{\sin \theta} \partial_{\theta} \sin \theta \ \partial_{\theta} \ f_{l}(\theta) = l(l+1) f_{l}(\theta),  
\label{eigenequation1} 
\ee
and satisfy an ortho-normality condition under the inner product defined by the 
$\theta, \ \varphi$ integral with a weight $\sqrt{-g}/R^{2}$,  
\bea 
\langle f_{l'}| f_{l} \rangle 
&=& \frac{1}{R^{2}} \int_0^\pi d\theta \int_0^{2\pi} d\varphi \ \sqrt{-g} \ f_{l'}(\theta) f_{l}(\theta) \nonumber \\ 
&=& 
 \int_0^\pi \sin \theta \ d\theta \int_0^{2\pi} d\varphi \ f_{l'}(\theta) f_{l}(\theta) = \delta_{ll'}. 
\label{zeromode2} 
\eea
The mode function for the ``super-partner" $A_{\theta}$ can be easily found, by looking at the mixing term in (\ref{zeromode1}), $(\partial_{\theta} A_{\mu} - \partial_{\mu} A_{\theta})(\partial_{\theta} A^{\mu} - \partial^{\mu} A_{\theta})$. 
If the Higgs mechanism is operative, $\partial_{\theta} A^{\mu}$ and $\partial^{\mu} A_{\theta}$ should have the same mode function. Thus we expect the mode function $g_{l}(\theta)$ of $A_{\theta}$, 
\be 
A_{\theta} = \sum_{l=0}^{\infty} A_{\theta}^{(l)}(x) g_{l}(\theta),  
\ee
behaves as $g_{l}(\theta) \propto \partial_{\theta} f_{l}(\theta)$. By taking an inner products of the both sides of (\ref{eigenequation1}) with $f_{l'}(\theta)$, and using the orthonormality (\ref{zeromode2}), we get a relation 
\be 
\langle \partial_{\theta} f_{l'} | \partial_{\theta} f_{l} \rangle = \delta_{l,l'} l(l+1).  
\ee
Thus we easily see that correctly normalized $g_{l}$ are given by  
\be 
g_{l}(\theta) = \frac{1}{\sqrt{l(l+1)}} \partial_{\theta} f_{l}(\theta) \ \ (l \neq 0), 
\ee
with an orthonormality condition 
\be 
\langle g_{l'} | g_{l} \rangle = \delta_{ll'}. 
\ee
Differentiating (\ref{eigenequation1}) by $\theta$, the eigenvalue equation satisfied by $g_{l}$ is easily known to be  
\be 
- \partial_{\theta} \frac{1}{\sin \theta} \partial_{\theta} \ \sin \theta \ g_{l}(\theta) = l(l+1) g_{l}(\theta).  
\label{zeromode5}  
\ee 

Putting the differential operators appearing in the eigenvalue equations for $f_{l}$ and $g_{l}$  together, we get a Hamiltonian $H$ of a supersymmetric quantum mechanics, in the space of $(f_{l}, \ g_{l})^{t}$, which can be written in terms of two supercharges $Q_{1}$ and $Q_{2}$ of $N = 2$ supersymmetry as 
\bea 
H &=& 
\pmatrix{
 - \frac{1}{\sin \theta} \partial_{\theta} \sin \theta \partial_{\theta} & 0 \cr 
0   &  - \partial_{\theta} \frac{1}{\sin \theta} \partial_{\theta} \sin \theta \cr 
},  \nonumber \\ 
H &=& Q_{1}^{2} = Q_{2}^{2}, \nonumber \\  
Q_{1} &=&  
\pmatrix{
0 &  - \frac{1}{\sin \theta} \partial_{\theta} \sin \theta  \cr 
\partial_{\theta}   &  0 \cr 
}, \ \ 
Q_{2} =   
\pmatrix{
0 &  i \frac{1}{\sin \theta} \partial_{\theta} \sin \theta  \cr 
i \partial_{\theta}   &  0 \cr 
}, \ \ 
\{Q_{1}, Q_{2} \} = 0.  
\eea
Thus we know that the non-zero modes $(f_{l}, g_{l})$  or $(R A_{\mu}, A_{\theta})$ form a super-multiplet, whose infinitesimal super-transformation is governed by $Q_{1,2}$. It is easy to see that the 4D mass-squared operators for $(R A_{\mu}, A_{\theta})$, obtained by a direct calculation of the action (\ref{zeromode1}) by putting a suitable gauge fixing term in order to eliminate the mixing terms between $A_{\mu}$ and $A_{\theta}, \ A_{\varphi}$, just
coincide with the operators in $H$. The gauge-fixed action tells us that the differential operator for 4D mass-squared of the remaining field $\tilde{A}_{\varphi} \equiv \frac{A_{\varphi}}{\sin \theta}$, which may be understood as the field obtained by multiplying zweibein $e_{2} \ ^{\varphi}$ to $A_{\varphi}$, is exactly the same as that for $g_{l}$. We thus learn $\tilde{A}_{\varphi}$ can be expanded as 
\be  
\tilde{A}_{\varphi} = \sum_{l=0}^{\infty} A_{\varphi}^{(l)}(x) g_{l}(\theta),  
\ee
just as for $A_{\theta}$. 

Substituting these mode expansions for $A_{\mu}, \ A_{\theta}, \ \tilde{A}_{\varphi}$ in the lagrangian (\ref{zeromode1}), after some arithmetic by use of $\partial_{\theta} f_{l}(\theta) = \sqrt{l(l+1)} g_{l}(\theta)$, eigenvalue equations and ortho-normality conditions, we arrive at, after $\theta, \ \varphi$ integrations,  
 the following simple expression of the 4D effective lagrangian for the gauge-Higgs sector. (Here we ignore the zero-modes for $A_{\theta}$ and $\tilde{A}_{\varphi}$, which will be separately discussed below)  
\bea 
{\cal L}_{eff}^{4D} &=& 
\frac{1}{4}\sum_{l=0}^{\infty} F_{\mu \nu}^{(l)} F^{(l)\mu \nu}  \nonumber \\ 
&-& \frac{1}{2}\sum_{l=1}^{\infty} [\partial_{\mu} A^{(l)}_{\theta}(x) - 
\frac{\sqrt{l(l+1)}}{R} A_{\mu}^{(l)}(x)][\partial^{\mu} A^{(l)}_{\theta}(x) - \frac{\sqrt{l(l+1)}}{R} A^{(l)\mu}(x)] \nonumber \\ 
&-& \frac{1}{2}\sum_{l=1}^{\infty} (\partial_{\mu} A^{(l)}_{\varphi}(x))(\partial^{\mu} A^{(l)}_{\varphi}(x)) 
- \frac{1}{2R^{2}} \sum_{l=1}^{\infty} l(l+1) (A^{(l)}_{\varphi}(x))^{2}. 
\label{zeromode3} 
\eea
This lagrangian clearly shows that a Higgs-like mechanism is operative for the 
each sector of non-zero modes of $(A^{(l)}_{\mu}(x), A^{(l)}_{\theta}(x))$ system, with $A^{(l)}_{\theta}(x)$ behaving as a would-be N-G boson, while $A^{(l)}_{\varphi}(x)$ remain as physical fields with masses $\frac{l(l+1)}{R^{2}}$. In fact, the second line of (\ref{zeromode3}) is nothing but the lagrangian for a non-linear sigma model of $A^{(l)}_{\theta}(x)$ with a ``vacuum expectation value" $\frac{\sqrt{l(l+1)}}{R}$.   

Now we will consider the zero-modes for $A_{\theta}$ and $\tilde{A}_{\varphi}$, namely the eigenfunction with a vanishing 4D mass, $g_{0}(\theta)$. From the 
differential equation (\ref{zeromode5}), it is easy to find the general solution for $g_{0}$, 
\be 
g_{0}(\theta) = c_{1} \frac{1}{\sin \theta} + c_{2} \cot \theta \ \ \ (c_{1,2}:  \mbox{constants}). 
\label{monopole1}
\ee
We find the 4D kinetic terms for the 4D fields $A^{(0)}_{\theta}(x), \ A^{(0)}_{\varphi}(x)$, the potential candidates for our Higgs field, accompanied by this $g_{0}(\theta)$, are not normalizable, as 
\be 
\langle g_{0} | g_{0} \rangle = 2\pi \int_{0}^{\pi} \ \sin \theta \ (g_{0}(\theta))^{2} \ d\theta = \infty. 
\ee
We thus have to conclude the possible lightest 4D scalars $A^{(0)}_{\theta}(x), \ A^{(0)}_{\varphi}(x)$ disappear from the spectrum of our theory. 

One possible way out of this problem is to identify the next lightest physical scalar $A_{\varphi}^{(1)}(x)$, having a 4D mass-squared $\frac{2}{R^{2}}$, with our Higgs field in the low energy world. 
From the viewpoint to seek the solution of the hierarchy problem, then the compactification mass scale $M_{c} \equiv \frac{1}{R}$ should be of ${\cal O}(1 TeV)$, or so. 

If we are going to insist in a small extra dimension, such as $M_{c} = {\cal O} (M_{pl})$, we have to device some still unknown mechanism to get a normalizable zero-mode. Right now we have no concrete idea concerning this possibility. It may be worthwhile to study the possibility to include the higher powers or the derivatives of $F_{MN}F^{MN}$ 
in the original lagrangian, which modify the eigenvalue equations.  

We have learned in the 5D gauge-Higgs unification with $S^{1}$ as the extra space that A-B effect plays an important role 
for the Higgs mass. Thus, it may also worthwhile to think about some 
gauge field configuration of $A_{\theta}, \ \tilde{A}_{\varphi}$ representing 
some non-trivial magnetic property, such as a monopole configuration. 

From this viewpoint, it is interesting to note that when $\tilde{A}_{\varphi}$ takes one of the zero mode solutions  $g_{0}(\theta)$ in (\ref{monopole1}), $\frac{1 - \cos \theta}{\sin \theta} \ \ (c_{1} = -c_{2} = 1)$, with $A_{\theta} = 0$, the field configuration just describes a magnetic monopole at the origin of $S^{2}$ 
compensated by a Dirac string penetrating through the south pole and ending at the origin of $S^{2}$. Another independent solution, $\frac{1 + \cos \theta}{\sin \theta}$ corresponds to a Dirac string penetrating through the north pole. Thus the Higgs zero modes really have close relation with the monopole.

\section{Summary}
In this paper, we have calculated the one-loop corrections to the 
mass of Higgs identified with the extra components of the gauge field 
in a six dimensional massive scalar QED compactified on the two-sphere $S^{2}$.  
We have explicitly shown that the radiatively induced Higgs mass is finite in general, i.e. for an arbitrary bulk scalar mass $M$. 
Furthermore, we have demonstrated that the remaining finite part actually vanishes, 
at least for small bulk mass $M$. Though we have not explicitly shown, these results strongly suggest that the radiatively induced Higgs mass exactly vanishes for arbitrary bulk mass. This conclusion may have a simple physical interpretation 
that in the simply-connected space like $S^{2}$ the Wilson-loop, which is non-trivial in the case of non-simply-connected extra space such as $S^{1}$ and is 
responsible for the finite Higgs mass at quantum level, can be always shrunk to a point and therefore the finite Higgs mass is not generated at quantum level. 

Concerning the technical aspect of our one-loop calculation, the Kaluza-Klein mode sum  before the 4D momentum integration, which was possible in the case of odd-dimensional sphere like $S^{1}$, was not possible. We thus first performed the 4D momentum integration by use of dimensional regularization method, and wrote the result in terms of the product of a gamma function with a series of zeta functions. In the process a careful treatment of the single pole appearing in the zeta function at $z=1$ was necessary before taking the limit $\epsilon \to 0 \ (\epsilon = \frac{d-4}{2})$. We have checked, though it has not been shown in this paper, that the same technique can be applied also for the case of one-loop calculation in 5D gauge-Higgs unification on $S^{1}$. 

In order to identify the Higgs field with some Kaluza-Klein mode of the extra 
space components of the gauge fields $A_{\theta}, \ A_{\varphi}$, we investigated the mass spectrum and corresponding mode functions. We have shown that the each sector of non-zero Kaluza-Klein modes has a Higgs-like mechanism and has a $N=2$ quantum mechanical supersymmetry, in accordance with the general argument in \cite{LNSS}. It was also shown that, although a suitable linear combination of $A_{\theta}, \ A_{\varphi}$ is absorbed as a would-be Nambu-Goldstone boson to $A_{\mu}$, there remains an independent massive physical scalar, in each sector. 

The kinetic terms of zero-modes of $A_{\theta}, \ \tilde{A}_{\varphi} \equiv \frac{A_{\varphi}}{\sin \theta}$, which are supersymmetric states and the possible candidates of the Higgs, were demonstrated to 
be non-normalizable, and therefore these zero-modes disappear from the spectrum of the theory.  We thus argued how we should interpret this result 
in the context of the hierarchy problem, depending on the supposed order of 
the compactification scale $M_{c} = 1/R$.  

In this context, an interesting claim was made that the zero-mode solutions of 
the eigenvalue equations to fix the 4D mass-squared just correspond to the gauge 
field configuration describing a magnetic monopole at the origin of $S^{2}$ 
accompanied by a Dirac string. Let us recall that also in the 5D gauge-Higgs unification scenario, the zero mode of 4D scalar and its radiatively induced finite mass has a close relation with a Wilson-loop, or A-B effect, which is also due to 
a non-trivial gauge field configuration describing magnetic flux. 
 
There remain many issues worth while studying. For instance, it would be interesting to extend our analysis in the present paper to 
more general case with higher dimensional spheres, the gauge-Higgs unification 
on $M^4 \times S^N \ (N \geq 3)$, to see if some new approach is available from  the viewpoint of the hierarchy problem. We naively expect that $S^{N} \ (N \geq 3)$ are also simply-connected spaces and therefore the radiatively induced Higgs mass exactly vanishes in the generalized case too. An explicit demonstration, however, will be necessary to confirm if this really is the case. 

In this paper we calculated the two point function of $A_{\theta}$. This corresponds to the calculation of the second derivative at the origin of the 
effective potential of $A_{\theta}, \ A_{\varphi}$. Thus it may be also interesting to calculate the effective potential itself and take the derivative at the minimum, or equivalently to calculate the two point function under the presence of non-zero background fields of $A_{\theta}, \ A_{\varphi}$, though we are not sure whether such effective potential is ever induced at quantum level (It might happen the effective potential just vanishes again due to the trivial Wilson loop).

We finally briefly comment on the radiative correction to the Higgs mass in the 6D Gravity-Gauge-Higgs unification model on $M^{4} \times S^{2}$. According to our preliminary result, in this case the radiatively induced Higgs mass is logarithmically divergent unlike the case of the present paper and the cases of 5D gauge-Higgs unification \cite{HIL} and 5D Gravity-Gauge-Higgs unification \cite{HLM} models. 
This divergence seems to have its origin in the fact that the Casimir energy is  
logarithmically divergent for the even dimensional spheres \cite{Myers}. 

\vspace*{1cm}
\begin{center}
{\bf Acknowledgements}
\end{center}  
The work of C.S.L. was supported by the Grant-in-Aid for Scientific Research 
of the Ministry of Education, Science and Culture, No.15340078, No. 18204024. 
The work of N.M. was supported by INFN, sezione di Roma. 
The work of K.H. was supported by the Science and Engineering Research,  
Canada. 
\begin{appendix}
 
\setcounter{equation}{0}
\section{Derivation of 4D effective action of the bulk scalar field}
\label{derive}
In this appendix, we describe in detail 
the derivation of the 4D effective action for the bulk scalar field. 
The starting point is the second line in (\ref{4Deff})
\bea
&&R^2 \int_0^\pi d\theta 
\int_0^{2\pi} d\varphi \sin \theta 
\left[
\Phi^* \left( -\partial_\mu \partial^\mu + \frac{1}{R^2} \Delta_{S^2} 
- M^2 \right)\Phi \right. \nonumber \\
&& \left. -i\frac{ce}{R^2}
\left(
\Phi^* \sin \theta \partial_\theta \Phi 
-\Phi \sin \theta \partial_\theta \Phi^*
\right) 
-\frac{c^2 e^2}{R^2}\Phi^* \sin^2 \theta \Phi
\right] \\ 
\label{eff}
&=& R^2 \int_0^\pi d\theta 
\int_0^{2\pi} d\varphi \sin \theta \sum_{l',m'} \sum_{l,m}
\left[
Y_{l'}^{m'}(\theta, \varphi)^* 
\left( -\partial_\mu \partial^\mu + \frac{1}{R^2} \Delta_{S^2} 
- M^2 \right)Y_l^m(\theta, \varphi) \right. \nonumber \\
&& \left. -i\frac{ce}{R^2}
\left(
Y_{l'}^{m'}(\theta, \varphi)^* 
 \sin \theta \partial_\theta Y_l^m(\theta, \varphi)
-Y_l^m(\theta, \varphi)
\sin \theta \partial_\theta 
Y_{l'}^{m'}(\theta, \varphi)^* 
\right) \right. \nonumber \\
&& \left. 
-\frac{c^2 e^2}{R^2} Y_{l'}^{m'}(\theta, \varphi)^* 
\sin^2 \theta Y_l^m(\theta, \varphi)
\right] \phi _{l',m'}(x)^* \phi_{l,m}(x) 
\label{4Deff1}
\eea
where the mode expansion in terms of spherical harmonics $Y_l^m(\theta, \varphi)$,  
\bea 
\Phi (x, \theta, \varphi) &=& \sum_{l=0}^{\infty} \sum_{m=-l}^{l} \ \phi_{l,m}(x) Y_{l}^{m}(\theta,\varphi),  \nonumber \\ 
Y_l^m(\theta, \varphi) &=& (-1)^m \sqrt{\frac{2l+1}{4\pi}\frac{(l-m)!}{(l+m)!}} 
P_l^m(\cos \theta) e^{i m \varphi}
\label{spherical}
\eea
where $P_l^m(\cos\theta)$ being Legendre polynomials, 
is substituted in the second line. 
The first line of the r.h.s. of (\ref{4Deff1}) is trivial since we know that eigenvalues of $\Delta_{S^2}$ is given by $-l(l+1)$.  
What are nontrivial in the calculation are 
the $\theta, \varphi$ integrations in the second and the third lines.  

Let us first calculate the integral
\bea
\int_0^\pi d\theta \int_0^{2\pi} d \varphi 
\sin \theta Y_{l'}^{m'}(\theta, \varphi)^* 
 \sin \theta \partial_\theta Y_l^m(\theta, \varphi). 
\label{1st}
\eea
We note the relation
\bea
\sin \theta \frac{dP_l^m(\cos \theta)}{d\theta} 
= \frac{1}{2l+1}
\left[
l(l-m+1)P_{l+1}^m(\cos \theta) 
-(l+1)(l+m) P_{l-1}^m(\cos \theta)
\right]
\label{Legendre1}
\eea
derived from the formulae 
\bea
\sin \theta \frac{dP_l^m (\cos \theta)}{d\theta} 
&=& -(l+1)\cos \theta P_l^m(\cos \theta) +(l-m+1)P_{l+1}^m (\cos \theta), \\
\cos\theta P_l^m(\cos \theta) 
&=& \frac{l-m+1}{2l+1}P_{l+1}^m(\cos \theta) 
+ \frac{l+m}{2l+1}P_{l-1}^m(\cos \theta). 
\eea
Putting (\ref{spherical}) and (\ref{Legendre1}) into (\ref{1st}), 
we obtain
\bea
&&\int_0^\pi d\theta \int_0^{2\pi} d \varphi 
\sin \theta Y_{l'}^{m'}(\theta, \varphi)^* 
 \sin \theta \partial_\theta Y_l^m(\theta, \varphi) 
 \nonumber \\
&=& \delta_{m',m} \{ 
\delta_{l',l+1} \frac{l}{\sqrt{(2l+1)(2l+3)}}\sqrt{(l+1)^{2}-m^{2}} 
- \delta_{l',l-1} \frac{l+1}{\sqrt{(2l-1)(2l+1)}}\sqrt{l^{2}-m^{2}} 
\}, \nonumber \\ 
\label{relation} 
\eea
where we used the orthonormal condition for the Legendre Polynomials
\bea
\int_0^\pi d\theta \sin \theta P_{l'}^m(\cos\theta) P_l^m(\cos \theta) 
= \delta_{ll'}\frac{2(l+m)!}{(l-m)!(2l+1)}. 
\label{orthonormal}
\eea
Thus, we obtain the three point vertices in 4D effective action, 
\bea
&&-ice\int_0^\pi d\theta \int_0^{2\pi} d\varphi \sin \theta 
(\Phi^* \sin \theta \partial_\theta \Phi 
- \Phi \sin \theta \partial_\theta \Phi^*) 
\nonumber \\
&=& -2ice \sum_{l,m} \sqrt{\frac{(l+1)^2-m^2}{(2l+1)(2l+3)}}(l+1)
\left[
\phi_{l+1,m}(x)^* \phi_{l,m}(x)
- \phi_{l,m}(x)^* \phi_{l+1,m}(x)
\right]. 
\label{3ptv}
\eea
Next nontrivial integral is 
\bea
\int_0^\pi d\theta \int_0^{2\pi} d \varphi 
\sin \theta Y_{l'}^{m'}(\theta, \varphi)^* 
 \sin^{2} \theta Y_l^m(\theta, \varphi). 
\label{2nd}
\eea
In this case, using (\ref{spherical}), (\ref{orthonormal}) 
and the following formulae twice 
\bea
P_{l-1}^{m+1}(\cos\theta) - P_{l+1}^{m+1}(\cos \theta) 
= -(2l+1)\sin \theta P_l^m(\cos \theta),
\eea
we obtain 
\bea
&&\int_0^\pi d\theta \int_0^{2\pi} d \varphi 
\sin \theta Y_{l'}^{m'}(\theta, \varphi)^* 
 \sin^{2} \theta Y_l^m(\theta, \varphi) 
 \nonumber \\
&=& \delta_{m',m}
\left[
\delta_{l',l}\frac{2(l^2+l-1+m^2)}{(2l-1)(2l+3)} 
-\delta_{l',l-2}\frac{\sqrt{(l^2-m^2)((l-1)^2-m^2)}}{(2l-1)
\sqrt{(2l-3)(2l+1)}} \right. \nonumber \\
&& \left. -\delta_{l',l+2}\frac{\sqrt{((l+2)^2-m^2)((l+1)^2-m^2)}}{(2l+3)
\sqrt{(2l+1)(2l+5)}} 
\right]. 
\eea
This leads to the following four point vertex in 4D effective action,
\bea
&&-c^2 e^2 \int_0^\pi d\theta \int_0^{2\pi} d\varphi 
\sin \theta \Phi^* \sin^2 \theta \Phi \nonumber \\
&=& -c^2 e^2 \sum_{l,m}
\left[
\frac{2(l^2+l-1+m^2)}{(2l-1)(2l+3)} \phi_{l,m}(x)^*\phi_{l,m}(x) 
\right. \nonumber \\
&& \left. 
-\frac{\sqrt{((l+2)^2-m^2)((l+1)^2-m^2)}}{(2l+3)\sqrt{(2l+1)(2l+5)}} 
(\phi_{l,m}(x)^* \phi_{l+2,m}(x) + \phi_{l+2,m}(x)^* \phi_{l,m}(x))
\right]. \nonumber \\
\label{4ptv}
\eea
Putting (\ref{3ptv}) and (\ref{4ptv}) into (\ref{eff}), 
the final result (\ref{4Deff}) is obtained.

\setcounter{equation}{0}
\section{Correct treatment of the zeta function}
\label{zeta}
In this appendix, in order to show the validity of the method we adopted in 
the text concerning the treatment of the zeta function 
$\zeta(z)=\sum_{l=1}^\infty 1/l^z$, i.e. first $z$ is shifted from 
an integer and at the final stage of calculation the limit of $z$ approaching to the integer is taken, we consider the following simple summation. 
\bea
\sum_{l=0}^\infty(2l+1). 
\eea
If we naively compute this summation as a difference of zeta functions, we obtain
\bea
\sum_{l=0}^\infty (2l+1) = \sum_{l=0}^\infty [(l+1)^2-l^2] 
= \sum_{l=0}^\infty (l+1)^2-\sum_{l=0}^\infty l^2 
= \zeta(-2)-\zeta(-2) =0. 
\eea
On the other hand, this result seems to contradicts with the following result by direct replacement to the zeta function.  
\bea
\sum_{l=0}^\infty (2l+1) = 2\sum_{l=0}^\infty l + \sum_{l=0}^\infty 1 
= 2\zeta(-1) +1 + \zeta(0) = 2\left( -\frac{1}{12} \right) + 1 + 
\left(-\frac{1}{2} \right) = \frac{1}{3}. 
\label{contra}
\eea
This superficial contradiction should come from the naive treatment of the sum or difference of divergent quantities. So to resolve the problem we make the zeta function $\zeta(z)$ analytically continued to a noninteger $z$ and later take the limit $z \to {\rm integer}$. Namely 
\bea
\zeta(-2)-\zeta(-2) &=& \lim_{\epsilon \to 0} 
\left[
\sum_{l=0}^\infty(l+1)^{2+\epsilon} - \sum_{l=0}^\infty l^{2+\epsilon}
\right] \\
&=& \lim_{\epsilon \to 0} 
\left[1+
\sum_{l=1}^\infty \left\{ 
(l^{2}+2l+1)l^\epsilon \left( 1+\frac{1}{l} \right)^\epsilon \right\} 
- \sum_{l=1}^\infty l^{2+\epsilon}
\right] \\
&=& \lim_{\epsilon \to 0} 
\left[1+
\sum_{l=1}^\infty \left\{ 
l^{2+\epsilon} +2l^{1+\epsilon} + l^\epsilon 
+ \epsilon l^{\epsilon-1} + \epsilon(\epsilon-1)l^{\epsilon-1} 
\right. \right. \nonumber \\
&& \left. \left. + \frac{\epsilon(\epsilon-1)(\epsilon-2)}{6}l^{\epsilon-1} 
\right\} 
- \sum_{l=1}^\infty l^{2+\epsilon}
\right] \\
&=& \lim_{\epsilon \to 0} \left[ 1+
2\zeta(-1-\epsilon) + \zeta(-\epsilon) 
+ \epsilon \left( 1 + (\epsilon-1) \right. \right. 
\nonumber \\
&& \left. \left. + \frac{(\epsilon-1)(\epsilon-2)}{6}\right) \zeta(1-\epsilon) \right]. 
\label{correct}
\eea
We note that the last term in (\ref{correct}), though it is accompanied by $\epsilon$, has nonzero contribution due to 
the pole of the first order, present in $\zeta(1-\epsilon)$. 
That is why there appears the discrepancy 
between the results (\ref{contra}) and (\ref{correct}),  
even in the limit $\epsilon \to 0$. 
We thus learn that the naive calculation setting $\epsilon=0$ 
from the beginning leads to a wrong result. 
In fact, using 
\bea
\lim_{\epsilon \to 0}\epsilon\zeta(1-\epsilon) 
= \lim_{\epsilon \to 0} \epsilon \frac{1}{-\epsilon} =-1, 
\eea
(\ref{correct}) becomes
\bea
1+
2\zeta(-1) + \zeta(0) 
- \left( 1 + (-1) 
+ \frac{(-1)(-2)}{6} \right)= \frac{1}{3}-\frac{1}{3} =0, 
\eea
which is consistent with the fact $\zeta(-2)-\zeta(-2)=0$. 

\setcounter{equation}{0}
\section{A formula for the summation of zeta functions} 
\label{formula}  
In this appendix, we derive a mathematical formula, which is useful in 
the process to show the vanishing Higgs mass-squared $m_{H}^{2}$ for the case of 
massless scalar (see (\ref{masslessfinite}), (\ref{masslessfinite2})). The method shown below is easily applied for other summations, such as the one shown in (\ref{formula2}). 

The summation we consider is 
\bea 
\sum_{n=1}^\infty \frac{\zeta(2n+1)}{(2n+3)(2n+4)(2n+5)}.  
\eea
First we decompose zeta functions as 
\bea 
\zeta (2n+1) = S(2n+1) + 1, \ \ S(z) \equiv \sum_{l=2}^{\infty} \ \frac{1}{l^{z}} .  
\eea
Then by using a formula 
\bea 
\sum_{n=1}^{\infty} \ \frac{1}{(2n+3)(2n+4)(2n+5)} = \ln 2 - \frac{41}{60} , 
\eea
the summation 
\bea  
\sum_{n=1}^\infty \frac{\zeta(2n+1)}{(2n+3)(2n+4)(2n+5)} 
&=& \sum_{n=1}^\infty \frac{S(2n+1)}{(2n+3)(2n+4)(2n+5)} + \ln 2 - \frac{41}{60}. 
\label{appendixC1}  
\eea

The integral representation of zeta functions leads to 
\bea 
S(z) 
&=& 
\frac{1}{\Gamma (z)} \int_{0}^{\infty} \frac{e^{-t} t^{z-1}}{e^{t}-1} \ dt.  
\eea
Thus 
\bea 
&& \sum_{n=1}^\infty \frac{S(2n+1)}{(2n+3)(2n+4)(2n+5)} \nonumber \\  
&& =  
\int_{0}^{\infty} \frac{e^{-t}}{e^{t}-1} \ dt \ 
\sum_{n=1}^{\infty} \ \frac{t^{2n}}{(2n)!(2n+3)(2n+4)(2n+5)}.  
\eea
We learn this summation can be written as (one may perform indefinite integral of $t^{2}(\cosh t -1)$ three times and then divide by $t^{5}$) 
\bea 
\sum_{n=1}^{\infty} \ \frac{t^{2n}}{(2n)!(2n+3)(2n+4)(2n+5)} 
&=& (\frac{1}{t^{3}} + \frac{12}{t^{5}}) \sinh t - \frac{6}{t^{4}}(\cosh t + 1) 
- \frac{1}{60},   
\eea
which leads to 
\bea 
&& \sum_{n=1}^\infty \frac{S(2n+1)}{(2n+3)(2n+4)(2n+5)} 
= \int_{0}^{\infty} \frac{e^{-t}}{e^{t}-1}\{ (\frac{1}{t^{3}} + \frac{12}{t^{5}}) \sinh t - \frac{6}{t^{4}}(\cosh t + 1) - \frac{1}{60} \} \ dt  \nonumber \\ 
&& = \int_{0}^{\infty} e^{-t} \{ (\frac{1}{t^{3}} + \frac{12}{t^{5}}) \frac{1+e^{-t}}{2} - \frac{6}{t^{4}}(\frac{1-e^{-t}}{2} + \frac{2}{e^{t}-1}) -\frac{1}{60} 
\frac{1}{e^{t}-1} \} \ dt . 
\eea 
Each term of the above equation is readily known to be written in terms of gamma function or the product of gamma and zeta functions. But, each term is divergent, though the whole expression should be finite. We thus introduce a factor $t^{\epsilon}$ to be multiplied by the integrand for the regularization, and take the limit $\epsilon \to 0$ in the final stage. Thus, after some arithmetic we arrive 
at 
\bea  
&& \sum_{n=1}^\infty \frac{S(2n+1)}{(2n+3)(2n+4)(2n+5)} \nonumber \\  
&& = \lim_{\epsilon \to 0} 
\{
\frac{2^{2-\epsilon}+1}{2} \Gamma (-2+\epsilon) + 3(2^{3-\epsilon}-1) \Gamma (-3+\epsilon) + 6(2^{4-\epsilon}+1) \Gamma (-4+\epsilon) \nonumber \\ 
&& \ \ \ \ \ \ \ \ -12 \Gamma (-3+\epsilon) (\zeta (-3+\epsilon) -1) - \frac{1}{60} \Gamma (1+\epsilon) (\zeta (1+\epsilon) -1)   
\}.  
\eea 
Noting 
\be 
\Gamma (\epsilon) = \frac{1}{\epsilon} - \gamma, \ \ 
\zeta (1+\epsilon) = \frac{1}{\epsilon} + \gamma, \ \ 
\Gamma (1+\epsilon) = \Gamma (1) + \Gamma'(1) \epsilon = 
\Gamma (1) (1+ \psi(1)\epsilon) = 1-\gamma \epsilon, \ \ 
\zeta (-3) = \frac{1}{120}, 
\ee
we find the pole disappears in the above equation and we get a finite result 
\be 
\sum_{n=1}^\infty \frac{S(2n+1)}{(2n+3)(2n+4)(2n+5)} 
= -\frac{\gamma}{60} - \ln 2 + \frac{499}{720} + 2 \zeta' (-3). 
\ee
From (\ref{appendixC1}) we finally obtain a formula 
\be 
\sum_{n=1}^\infty \frac{\zeta(2n+1)}{(2n+3)(2n+4)(2n+5)} 
= -\frac{\gamma}{60} + \frac{7}{720} + 2 \zeta' (-3). 
\ee

\end{appendix}


\end{document}